\documentstyle[12pt,aps,psfig]{revtex}
\draft
\tighten
\begin{document}

\title{The entrance dynamics of polymers into a gel}
\author{Walter Kob$^*$ and Leticia F. Cugliandolo$^{**}$ }
\address{
$^*$ 
Institute of Physics, Johannes Gutenberg-University, \\
Staudinger Weg 7, D-55099 Mainz, Germany
\\
$^{**}$
\it Laboratoire de Physique Th{\'e}orique de l'{\'E}cole Normale 
Sup{\'e}rieure, \\
24 rue Lhomond, 75231 Paris Cedex 05, France and \\
 Laboratoire de Physique Th{\'e}orique  et Hautes {\'E}nergies, Jussieu, \\
5 {\`e}me {\'e}tage,  Tour 24, 4 Place Jussieu, 75005 Paris France
}
\date{3. June, 2000}
\maketitle
\begin{abstract}
We use computer simulations to study the (driven) dynamics of a charged
polymer in a disordered medium, thus mimicking the setting used in gel
electrophoresis. In agreement with experiments, we find that inside the
gel the mobility of the polymer is only a weak function of its length
$N$. In contrast to this, the mean entrance time into the gel, is a very
strong function of $N$, $\langle \tau_e \rangle \propto N^{1.3}$, and does
not show any sign of saturation with increasing $N$. We rationalize this
effect by means of a simple model and propose an experimental setup that
should allow to increase the separation ability of gel electrophoresis
significantly.

\end{abstract}


Understanding the dynamics of macromolecules in disordered porous
media is important in many situations of practical interest such
as filtration, oil recovery, transport through membranes, and gel
electrophoresis
\cite{review_electro,general_macro_porous,review1,review2,review3,review4},
the problem we treat in this letter. Gel electrophoresis is a simple but
powerful technique used to separate polyelectrolyte strands with different
length. One of its main fields of application is molecular biology,
where it is used to sequence DNA.  In its standard setting, fragments of
various lengths are injected in a thin  layer of gel, composed by, e.g.,
agarose powder and a buffer solution. The latter imparts a charge to each
base-pair. If an external electric field is applied, the fragments migrate
in its direction with shorter strands moving faster than longer ones.
This, in principle, allows one to separate polymers with different
length $N$.  However, for large $N$ the velocity depends only weakly on
$N$ and therefore it is not feasible to use gel electrophoresis as an
efficient separation technique for long molecules.

In the usual experimental setup, the molecules are driven through the
interface between the buffer solution and the gel by the external
field. Very recently this process has called the attention of
experimentalists who tried to determine if it can be used as a method
to improve the efficiency of separation.

Using fluorescence microscopy, Oana {\it et al.} 
\cite{Oana} 
concluded
that DNA molecules stay at the interface for a time that decreases
with increasing pore size. Unfortunately, due to the difficulty with
this technique to investigate many such entrance events, it is hard to
extract quantitative information, such as the dependence of the delay
time on the size of the molecules or the applied field.  In addition,
the limited spatial resolution of $0.1\mu m$ prevents a detailed knowledge
of what the polyelectrolytes actually do at the interface.

Klep{\'a}rn\'{\i}k {\it et al.}
\cite{Kleparnik1,Kleparnik2} 
investigated the
injection process by measuring the time that a polymer needs to propagate
a macroscopic distance. From its extrapolation to vanishing distance,
they inferred that the molecules are stuck at the interface during an
interval  that increases with length  and speculated that it will grow
with decreasing field.  They attributed the delay to an electro-osmotic
flow in the opposite direction of the field which hinders some of the
molecules from entering the gel.

In the past, numerical simulations have proven to be a very
useful tool to understand the motion of polymers in disordered
media 
\cite{general_macro_porous}.
In the context of electrophoresis,
some of the most significant results are the ones by Deutsch {\it
et al.} 
\cite{pulsed-Deutsch,Deutsch,Deutsch2} 
who solved numerically
the Langevin dynamics of a driven chain moving in a two dimensional
random array of obstacles, allowing, however, the self-intersection
of the polymer. For rather strong fields, the numerical results
showed the existence of U-like conformations inside the bulk of
the gel, a result which was later confirmed experimentally by video
microscopy 
~\cite{hernias_micro}.

Here we study the details of the entrance process of charged polymers
into the gel  by means of simulations. In particular we analyze how the
time it takes the polymer to enter depends on its size and the strength of
the field. Since simulations are not restricted by any resolution limit,
they allow us to identify  the conformations of the molecule during
the entrance in the gel, which in turn will help us to understand the 
details of this process. By using the gathered information, we propose
a setup that should significantly improve the separation-efficiency in
constant field electrophoresis.

 The polymer is modeled by a chain of $N$ beads connected to each
other by an unbreakable spring 
\cite{Kremer}.
The interaction between
two monomers $i$ and $j$, that are a distance $r_{ij}$ apart,  is
\begin{eqnarray}
V(r_{ij}) &=&
4 \epsilon \, [
(\sigma/r_{ij})^{12}
-
(\sigma/r_{ij})^{6}
]
\; -
\; K_0 R^2_0/2 \; \ln [ 1- (r_{ij}/R_0)^2]
\,
(\delta_{j, i+1} + \delta_{j, i-1}),
\label{eq1}
\end{eqnarray}
where $\delta_{i,j}$ is the Kronecker $\delta$.  The first term is
strongly repulsive at short distances and represents the excluded
volume. For computational efficiency it is truncated and shifted at
the location of its minimum, $r_m=2^{1/6}\sigma$. The second term
represents the bond between neighboring monomers. In the following
we use reduced units and  measure length, energy and time  in units
of $\sigma$, $\epsilon$, and $\sqrt{\sigma m/\epsilon}$, where $m$
is the mass of one monomer and the Boltzmann constant $k_B$ is set
to 1. The constants $K_0$ and $R_0$ are 30 and 1.5, respectively. We
mimic the charging effect by assigning a charge $q=-1$ to each monomer.
(Note that we neglect the Coulomb interaction between monomers since
it can be assumed that the solvent leads to strong screening.)  If we
assume that the polymer is a DNA strand with charge $\sim 150 e \sim
150 \times 1.6 \, 10^{-19}$C per base pair, these numbers translate
into physical units of around 200{\AA}$\,$ for the unit of length and
2500 V/cm for the unit of field. The gel is modeled by a disordered
2d array of point particles obtained by equilibrating a system of soft disks
interacting via a $r^{-12}$ potential at $T=1$~\cite{footnote}. A snapshot of this system
was expanded by a factor of 5 and the resulting, {\it permanently fixed},
configuration with a liquid-like structure at short range, is our 2d
gel. The typical distance between neighboring particles is $\sim 7$
which gives pore sizes of the order of $\sim 5-15$. The size of the
gel perpendicular to the field is 160, and is sufficiently large to
avoid finite size effects if periodic boundary conditions are used.
The interaction of the polymer with the gel is given by a truncated
and shifted Lennard-Jones potential with the form of the first term in
Eq.~(\ref{eq1}). The solvent imparts a stochastic force on each monomer,
that is mimicked by substituting after each time step the velocities of
the monomers by new ones drawn from a Maxwell-distribution at $T=1$. The
positions of the monomers are then updated by means of the velocity
version of the Verlet algorithm, using a time-step $=0.01$.

At the beginning of each simulation we equilibrate the polymer by taking
into account only the monomer-monomer interactions. Typically, the
polymer acquires an open random-walk like conformation with a Kuhn length
$b \sim 2-5$ (estimated from visual inspection), which is of the order
of the pore size and corresponds to $400-1000${\AA}.  Once the polymer
is equilibrated we slide it in the direction of the field until it is
at a distance $r_m$ from the closest particle in the gel. This instant
defines the initial time $t=0$ when the polymer starts interacting with
the gel. We then turn on the field, driving the polymer into the gel.

We have studied polymer sizes $N=25$, 50, 100, 200, and 400 and
fields $E=0.0312$, 0.0625, 0.125, 0.25, 0.5, and 1.0.  In all runs the
temperature is such that $k_B T=1$, which we identify with room temperature
$T=300K$.  This leads to an adimensional parameter $qEb/k_B T$ between
$0.1$ and $10$. To improve the statistics we average all results over
at least 30 independent runs.

\begin{figure}[h] 
\psfig{figure=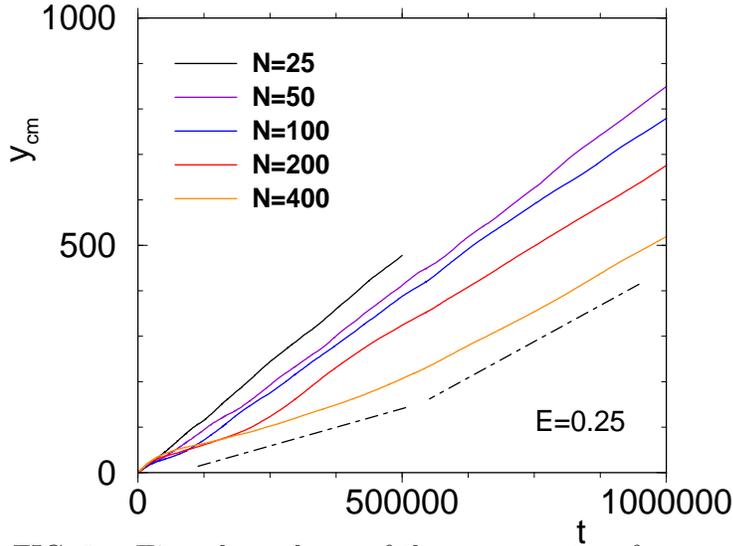,width=10cm,height=7.2cm}
\caption{
Time dependence of the mean center-of-mass position
in the direction of the field for
different polymer lengths. There are two time
regimes for the long polymers, indicated by the two dashed-dotted lines.
}
\label{fig1}
\end{figure}

\begin{figure}[h] 
\psfig{figure=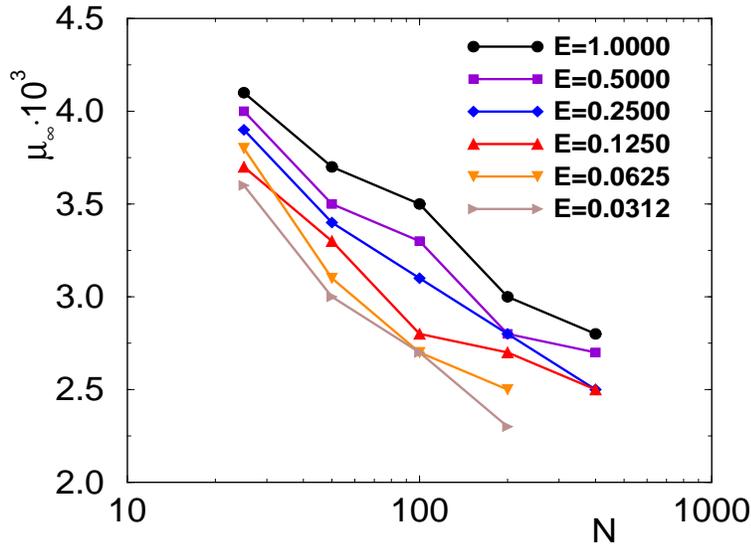,width=10cm,height=7.2cm}
\caption{ 
Size-dependence of the mobility at long times for different
external fields.
}
\label{fig2}
\end{figure}

One possibility to characterize the dynamics {\it inside} the gel
is to monitor the time dependence of the averaged center-of-mass
position in the direction of the applied field, $y_{cm}(t)$, shown
in Fig.~\ref{fig1} for various $N$.  After a transient, the curves
become straight lines with slope $v(t)$.  In Fig.~\ref{fig2} we plot
the mobility at large times, $\mu _{\infty}=\lim_{t \to \infty} v(t)/E$,
as a function of $N$. We see that to a good approximation $\mu _{\infty}$ is
independent of $E$ and that it is only a weak function of
$N$. (A change of $N$ by a factor of 100 changes $\mu _{\infty}$ only
by a factor of less than two.)  The reason for this is that for long
polymers not only the force due to the field is proportional to $N$
but also the effective friction  due to the gel is linear in $N$. For
large $N$ the two forces cancel  and the resulting velocity has only a
weak $N$-dependence 
\cite{review1,review2,review3}. 
Therefore it
is very hard to separate strands with similar lengths.

\begin{figure}[h] 
\psfig{figure=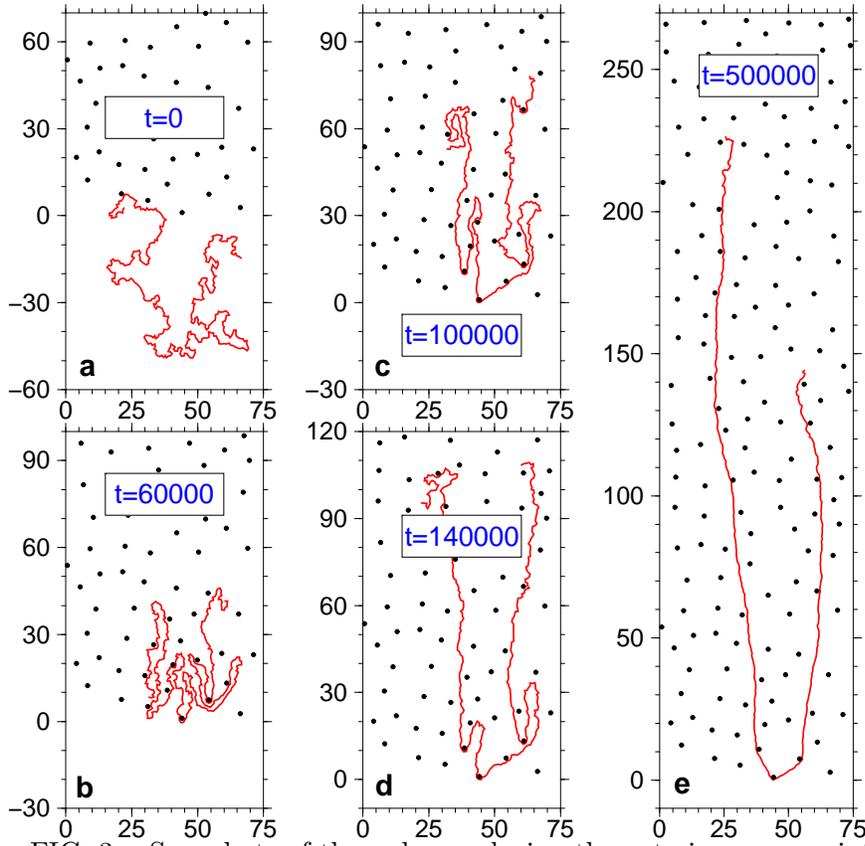,width=11.5cm,height=11cm}
\caption{ 
Snapshots of the polymer during the entering process into the gel.
The black circles represent the obstacles in the gel and their size
corresponds, approximately, to the range of the interaction. The external
field is pointing upwards. $N=400$ and $E=0.25$.
}
\label{fig3}
\end{figure}

{\mbox From} Fig.~\ref{fig1} we also recognize that at short times the mean
velocities have a stronger $N$-dependence than at long times, when
the polymer is in the bulk.  In Fig.~\ref{fig3} we show a series of
snapshots that help us understanding this result.  At early times,
Fig.~\ref{fig3}a, the field drives the (relatively open) equilibrium
configuration in an unhindered fall towards the gel. The polymer hits
the gel and its mobility decreases slightly. After a short interval
some monomers, either at the end or in the middle of the polymer, find
their way into the pores while others are still pinned at the boundary,
Fig.~\ref{fig3}b. The configuration acquires little hernias that
penetrate the gel, Fig.~\ref{fig3}c.  Subsequently, the little hernias
are eaten up by the longer ones
\cite{comment} 
and a ``staple-like''
configuration is reached with two elongated arms that penetrate the gel,
Fig.~\ref{fig3}d
\cite{comment2}.
For much longer times the longer arm
will finally drag the polymer into the gel. The transient dynamics due
to the entering process has ended and the asymptotic mean velocity
is reached. For large $N$, the time it takes to resolve this staple
configuration is determined by the difference in length between the
arms. Since this time is significantly longer that the time it takes the
polymer to reach this configuration, it determines the entrance delay
almost completely.

We mention that such staple-like configurations also occur {\it inside
the gel}, though with two important differences.  i) Unless the field is
extremely weak, their width is generally smaller, since the pinning is due
to only one or two obstacles. In the entering case instead, the pinning
occurs at more sites (therefore we call the configuration ``staple-like''
and not hernias) since outside the gel the polymer is open with a typical
cross-section that scales like the radius of gyration.  ii) Inside the
gel the generation of such pinned configurations is a {\it non-synchronized}
stochastic process.  At the interface most strands are trapped almost
immediately after touching the gel and hence the trapping is {\it
synchronized}.  These two points explain why the motion of the polymers
close to the surface is so much slower than inside the gel. We also
mention that none of the polymers we generated bounced back into the
solvent when they encountered the interface, i.e. moved against the
direction of the field. This shows that there can be a significant
delay also without an electro-osmotic flow, in contrast to the explanation
put forward by Klep{\'a}rn\'{\i}k {\it et al.}
\cite{Kleparnik1,Kleparnik2}
of their experimental results.

\begin{figure}[h] 
\psfig{figure=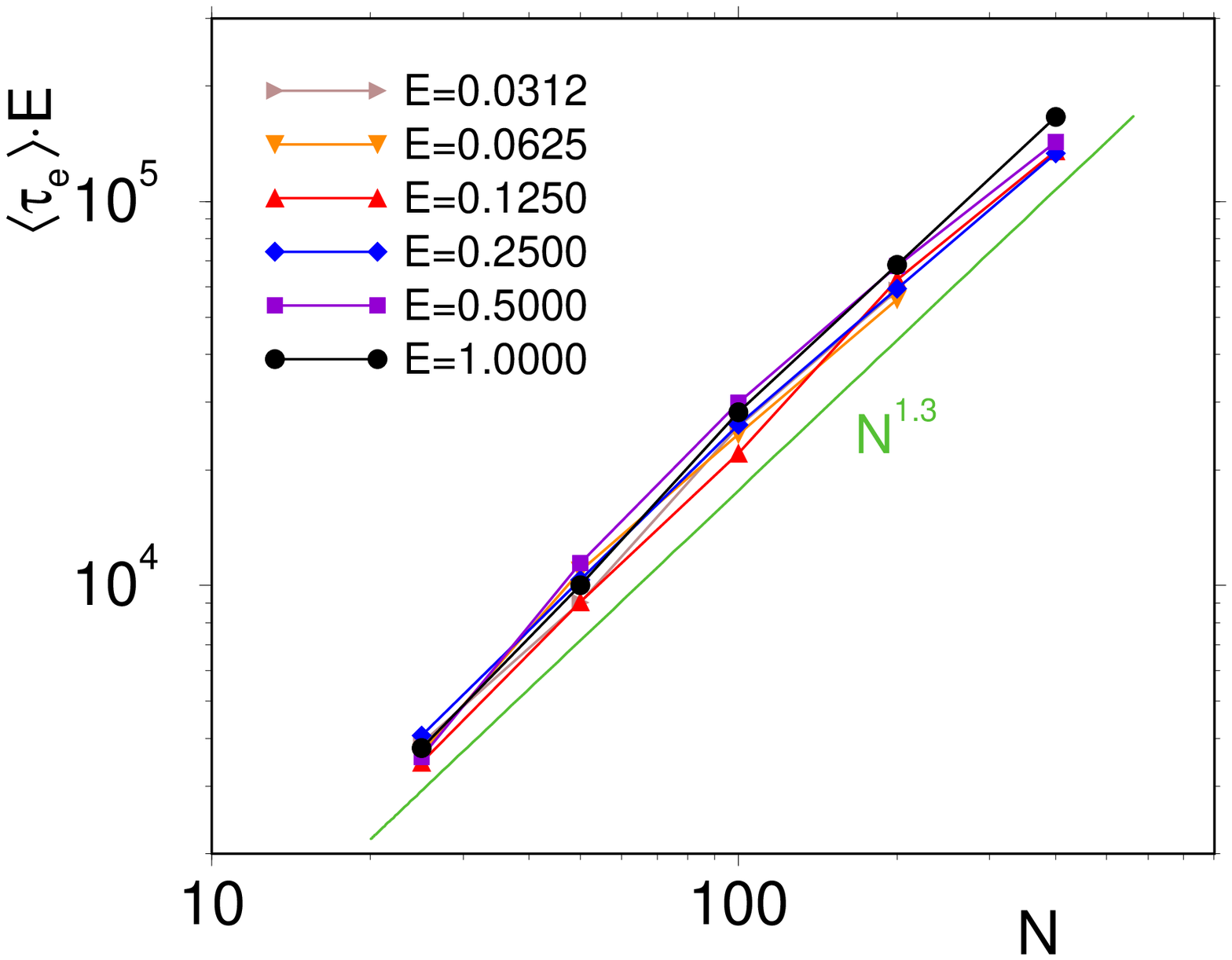,width=10cm,height=7.2cm}
\caption{ 
Scaled entering time $\langle \tau_e \rangle E$ as a function of the
polymer length for various values of the field $E$.
}
\label{fig4}
\end{figure}

In order to analyze how the entrance time depends on length and field,
we define a time $\tau_e$ such that $y_{cm}(\tau_e)=\alpha N$, with
$\alpha=0.75$. (The precise value of $\alpha$ is irrelevant as long as
the probability for the polymer to have entered more than $\alpha N$
without having been entangled is very small.) In Fig.~\ref{fig4} we show
$\langle \tau_e \rangle$ multiplied by $E$ versus $N$  for different
values of $E$. From this plot we conclude that the entering time is
inversely proportional to the field. Much more important, however, is the
strong dependence of $\langle \tau_e \rangle$ on $N$ (note the logarithmic
scale on the $\langle \tau_e \rangle$ axis!).  This dependence does not
saturate with $N$ and it is approximately given by the law $N^{1.3}$
(bold solid line). Therefore it should be possible to use this entering
process as a tool to increase the selectivity of gel electrophoresis.

A very simple model allows us to understand on a qualitative basis
the origin of this strong $N$-dependence: Let us assume, as suggested
by the simulations, that we can neglect the time needed to reach the
staple-like conformation and hence  can approximate $\tau_e$ by the time
necessary to resolve it. The problem can now be modelled by a massive
rope with uniform linear density $q$ that hangs under gravity, with
strength $E$, from a pulley
and whose motion is slowed down due to friction that is proportional
to its velocity. The arm carrying more mass  makes the rope slide in
its direction.  If we call $\delta$ the difference in length of the two
arms, normalized by the total length of the rope, it is easy to
calculate the time needed for
the rope to fall from the pulley and one finds

\begin{equation} 
\tau_e = - \frac{\xi N}{2qE} \; \ln\delta \quad ,
\label{eq2}
\end{equation} 
where $\xi$ is the friction constant per unit length.  In our problem
$\delta$ fluctuates from sample to sample.  Therefore an average over
samples translates into an average over $\delta$.  If we assume that the
distribution of $\delta$ is independent of $\delta$ we obtain $\langle
\tau_e \rangle \propto \xi N/E$. The stronger $N$-dependence shown in
Fig.~\ref{fig4} can be accounted for by assuming that the distribution
of $\delta$ has a somewhat larger weight at small $\delta$.

Equipped with the results of the simulation we propose an experimental
setup that should allow to increase the ability of gel electrophoresis
to separate polymers of different length.  In this setup we have an
alternating sequence of strips made out of gel and strips with just
solvent. If the field is pointing upwards, these strips are arranged
vertically  and have ``heights'' $g_1$, $s_1$, $g_2$, $s_2$, $g_3$, etc.
The basic idea is as follows.  If the polymer mixture is placed below the
first strip all polymers enter this strip, leading to the very strong
separation studied here. The polymers later leave the first strip and
enter a strip of solvent. The polymer changes its conformation from the
elongated structure acquired in the gel to a random walk-like structure
since in the pure solvent the elongated structure is not stable.
After some time it hits the next solvent-gel interface and the process
starts again. In order to allow a strong separability at the interface,
the width of the gel has to be at least on the order of the longest
polymers. To ensure that long polymers open up again during their fall in
the solvent, this strip also needs to be sufficiently long. Therefore it
is advisable to increase progressively the width of the strips i.e. $g_1<
g_2<g_3<\ldots$ and $s_1<s_2<s_3 \ldots$.

We conclude by stating that these results, together with the ones in
Refs.~\cite{Oana,Kleparnik1,Kleparnik2}, 
make us expect
that the use of interfaces can lead to a strong improvement of the
separability in gel electrophoresis.
 
We thank S. Bottani, M. Picco, and J.-L. Viovy for very useful
discussions.  Part of this work has been supported by the DFG through
grant SFB 262, the ITP in Santa Barbara, and the SPHINX collaboration
of the ESF.

\begin{thebibliography}{99}
\bibitem{footnote}
It can be expected that in this situation the dynamics of the polymer 
in two and three dimensions are qualitatively similar.
\bibitem{comment}
This is similar to the effect discussed by Deutsch 
\cite{pulsed-Deutsch}
and observed by Long and Viovy 
\cite{pulsed-Long} 
shortly after rotating
the field in pulsed field electrophoresis. However, the starting
equilibrium configuration is in our case a random walk-like structure
while in  pulsed-field electrophoresis the polymer has an elongated form.
\bibitem{comment2}
When all short hernias have been removed \cite{pulsed-Deutsch,pulsed-Long}
two possibilities arise: either the polymer enters though its two ends and
quickly acquires a ``staple-like'' conformation with two elongated arms,
or only one end and the middle section of the polymer enter, giving rise
to an elongated arm on one end and a double segment at the other end.
If this double arm has more mass that the elongated one, it will grow,
the polymer will acquire the staple-like configuration, and it will
eventually pull the polymer inside the gel.
\end{thebibliography}
\end{document}